\documentclass[submission,copyright,creativecommons]{eptcs}
\providecommand{\event}{TARK 2019} 
\usepackage{enumitem,amssymb}
\usepackage{amsmath}
\usepackage{breakurl}             
\usepackage{underscore}           

\newtheorem{proposition}{Proposition}

\newtheorem{definition}{Definition}

\title{Credible Information, Allowable Information \\and Belief Revision - Extended Abstract}
\author{
Giacomo Bonanno\thanks{I am grateful to three anonymous reviewers for helpful comments.}
\institute{University of California\\ Davis, USA}
\email{gfbonanno@ucdavis.edu}
}
\def\titlerunning{Credible and Allowable Information}
\def\authorrunning{G. Bonanno}

\begin{document}

\maketitle

\begin{abstract}
In an earlier paper [Rational choice and AGM belief revision, \emph{Artificial Intelligence}, 2009] a correspondence  was established between the choice structures of revealed-preference theory (developed in economics) and the syntactic belief revision functions of the AGM theory (developed in philosophy and computer science). In this paper we  extend the re-interpretation of (a generalized notion of) choice structure in terms of belief revision by adding: (1) the possibility that an item of ``information'' might be discarded as not credible (thus dropping the AGM success axiom) and (2) the possibility that an item of information, while not accepted as fully credible, may still be ``taken seriously'' (we call such items of information ``allowable''). We establish a correspondence between generalized choice structures (GCS) and AGM belief revision; furthermore, we provide a syntactic analysis of the proposed notion of belief revision, which we call \emph{filtered belief revision}.
\end{abstract}

\section{Introduction}
In \cite{Bon09} a correspondence was established between rational choice theory -- also known as revealed-preference theory\footnote{See, for example, \cite{Rot01} and \cite{Suz83}.} -- and the AGM theory of belief revision.\footnote{ \cite{AGM85,Gae88}}
\par
Revealed-preference theory considers \emph{choice structures} $\left\langle \Omega ,\mathcal{E} ,f\right\rangle$ consisting of a non-empty set $\Omega $ (whose elements are interpreted as possible alternatives to choose from), a collection \,$\mathcal{E}$ of subsets of $\Omega $ (interpreted as possible menus, or choice sets) and a function $f:\mathcal{E}\to 2^\Omega$   ($2^\Omega$ denotes the set of subsets of $\Omega $), representing choices made by the agent, conditional on each menu. Given this interpretation, the following restriction on the function $f$ is a natural requirement (the alternatives chosen from menu $E$ should be elements of $E$): $\forall E\in\mathcal E$,
\begin{equation}
\label{fEsubE}
f(E)\subseteq E.
\end{equation}
\par\noindent
The objective of reveled-preference theory is to characterize choice structures that can be ``rationalized" by a total
pre-order $\succsim$ on $\Omega$, interpreted as a preference relation,\footnote{
Thus the intended meaning of $\omega\succsim\omega'$ is ``alternative $\omega$ is considered to be at least as good as alternative $\omega'$".
}
in the sense that, for every $E\in\mathcal E$, $f(E)$ is the set of most preferred alternatives in $E$: $f(E)=\{\omega\in E: \omega \succsim \omega^\prime, \forall \omega^\prime\in E\}$.\smallskip
\par
The AGM theory of belief revision is a syntactic theory that takes as starting point a consistent and deductively closed set $K$ of formulas in a propositional language, interpreted as the agent's initial beliefs, and a function $B_{K}:\Phi \rightarrow 2^{\Phi }$  (where $\Phi $ denotes the set of formulas and  $2^{\Phi }$ the set of subsets of $\Phi $), called a \emph{belief revision function based on K}, that associates with every formula $\phi \in\Phi $ (interpreted as new information) a set $B_{K}(\phi )\subseteq \Phi $, representing the agent's revised beliefs in response to information $\phi$. If the function $B_{K}$ satisfies a set of six properties, known as the basic AGM postulates, then it is called a \emph{basic AGM belief revision function}, while if it satisfies two additional properties (the so-called supplementary postulates) then it is called a \emph{supplemented AGM belief revision function}. We denote a (basic or supplemented) AGM belief revision function  by $B^{\ast}_{K}$.\footnote{In the literature it is common to denote an AGM belief revision function by $\ast:\, \Phi \rightarrow 2^{\Phi }$ and to denote  by $K\ast\phi$ the belief set resulting from revising $K$ by $\phi$. However, we will continue to use the notation of \cite{Bon09}.}\smallskip
\par
In \cite{Bon09} the two theories were linked by means of a re-interpretation of the set-theoretic structures of revealed-reference theory, as follows. The set $\Omega $ is interpreted as a set of \emph{states}. A \emph{model based on} (or an \emph{interpretation of}) a choice structure $\left\langle \Omega ,\mathcal{E} ,f\right\rangle $  is obtained by adding a valuation $V$ that assigns to every atomic formula $p$ the set of states at which $p$ is true. Truth of an arbitrary formula at a state is then defined as usual. Given a model $\left\langle \Omega ,\mathcal{E},f,V\right\rangle $,  the \textit{initial beliefs} of the agent are taken to be the set of formulas $\phi $ such that $f(\Omega )\subseteq ||\phi ||$, where $||\phi ||$ denotes the truth set of $\phi $; hence $f(\Omega )$ is interpreted as the set of states that are initially considered possible. The events (sets of states) in $\mathcal{E}$ are interpreted as possible items of information. If $\phi $ is a formula such that $||\phi ||\in \mathcal{E}$, the \textit{revised belief upon learning that }$\phi $ is defined as the set of formulas $\psi $ such that $f(||\phi ||)\subseteq ||\psi ||$. Thus the event $f(||\phi ||)$ is interpreted as the set of states that are considered possible after learning that $\phi $ is the case. In light of this interpretation, condition (\ref{fEsubE}) above corresponds to the \emph{success postulate} of AGM theory (one of the six basic postulates): $\forall\phi\in\Phi$,
\begin{equation}
\label{successIntro}
\phi\in B_K(\phi),
\end{equation}
according to which any item of information is always accepted by the agent and incorporated into her revised beliefs.
\par
The correspondence between choice structures and AGM belief revision is as follows. First of all, define a choice frame to be \emph{strongly AGM-consistent} if, for every interpretation of it, the associated partial belief revision function (`partial' because, typically, there are formulas $\phi$ such that $||\phi||\notin\mathcal E$) can be extended to a (full-domain) supplemented AGM belief revision function (that is, one that satisfies the six basic AGM postulates as well as the two supplementary ones).
In \cite{Bon09} it is shown that a choice frame is strongly AGM-consistent if and only if it is ``rationalizable", that is, if and only if there is a total pre-order $\succsim$ on $\Omega$ such that, for every $E\in\mathcal E$, $f(E)=\{\omega\in E: \omega \succsim \omega^\prime, \forall \omega^\prime\in E\}$. In this context the interpretation of the relation $\succsim$ is no longer in terms of preference but in terms of plausibility: the intended meaning of $\omega\succsim\omega'$ is ``state $\omega$ is considered to be at least as plausible as state $\omega'$". Thus, for every item of information $E\in\mathcal E$, $f(E)$ is the set of most plausible states compatible with the information.\medskip
\par
In this paper we continue the analysis of the relationship between choice structures and AGM belief revision by removing restrictions \eqref{fEsubE} and \eqref{successIntro}, thus considering a more general notion of belief revision.\smallskip
\par
The success axiom has been criticized in the AGM literature on the grounds that individuals may not be prepared to accept every item of ``information'' as credible. For example, during the U.S. Presidential campaign in 2016, a "news" item appeared on several internet sites under the title ``"Pope Francis shocks world, endorses Donald Trump for president''.\footnote{
See: {\scriptsize \url{https://www.cnbc.com/2016/12/30/read-all-about-it-the-biggest-fake-news-stories-of-2016.html}
(accessed June 19, 2019).}
} While, perhaps, some people believed this claim, many discarded it as ``fake news''. In today's political climate, many items of ``information'' are routinely rejected as not credible.\smallskip
\par
There is a recent literature in the AGM tradition that relaxes the success axiom (\ref{successIntro}) and allows for some formulas to be treated as not credible, so that the corresponding ``information" is not allowed to affect one's beliefs (see \cite{FerHans99,Hansetal01,Mak97}). This paper's contribution follows this literature, while adding a further possibility.\smallskip
\par
First of all, we allow for some events -- in the set of potential items of information $\mathcal E$ -- to be treated as not credible, so that
\begin{equation}
\label{not credible}
f(E)=f(\Omega)\quad  \text{if } E\in\mathcal E \text{ \,is rejected as not credible}.
\end{equation}
Secondly, for information $E\in\mathcal E$ which \emph{is} credible we postulate the ``success'' property (\ref{fEsubE}):
\begin{equation}
\notag
f(E)\subseteq E,\quad  \text{if } E\in\mathcal E \text{ is credible}.
\end{equation}
 Finally, we also add a third type of information, which is \emph{taken seriously but not given the same status as credible information}. For example, a detective might have come to believe that of the three suspects suggested by preliminary evidence -- Ann, Bob and Carla -- Ann should be discarded in light of her impeccable past behavior, that is, the detective forms the belief that Ann is innocent. Suppose now that new evidence points to Ann as the person who committed the crime. In that case, while not forming the belief that Ann is indeed the culprit, the detective might now add Ann as a serious possibility, by no longer believing in her innocence; that is, the detective now \emph{considers it possible} that Ann is the culprit. We call an item of information that is taken seriously, while not treated as fully credible, \emph{allowable} and we capture possibility in terms of belief revision by the following condition that says that allowable information is not ruled out by the revised beliefs:
\begin{equation}
\label{plausible}
f(E)\cap E\neq\varnothing\quad \text{if } E\in\mathcal E \text{ \,allowable}.
\end{equation}
\par
We model credibility, allowability and rejection by partitioning the set $\mathcal E$ of possible items of information into three sets: the set $\mathcal E_C$ of \emph{credible} items, the set $\mathcal E_A$ of \emph{allowable} items and the set $\mathcal E_R$ of \emph{rejected} items. Thus we consider \emph{generalized choice structures} (GCS for short) $\left\langle \Omega ,\{\mathcal{E}_C, \mathcal E_A, \mathcal E_R\}, f\right\rangle$ such that:
\begin{enumerate}
  \item $\Omega\neq\varnothing$,
  \item $\mathcal{E}_C, \mathcal E_A, \mathcal E_R$ are mutually disjoint subsets of $2^\Omega$ with $\Omega\in\mathcal{E}_C$ and $\varnothing\in\mathcal{E}_R$,\footnote{These sets may be ``small", that is, we do not assume that $\mathcal{E}_C\cup \mathcal E_A\cup \mathcal E_R$ covers the entire set $2^\Omega$.
      }
  \item $f:\mathcal E\rightarrow 2^\Omega$ (where $\mathcal E=\mathcal{E}_C\cup \mathcal E_A\cup \mathcal E_R$) is such that
  \begin{enumerate}
    \item $f(\Omega)\neq\varnothing$,
    \item if $E\in\mathcal E_R$ then $f(E)=f(\Omega)$,
    \item if $E\in\mathcal E_C$ then $\varnothing\neq f(E)\subseteq E$,
    \item if $E\in\mathcal E_A$ then $f(E)\cap E\neq\varnothing$.
  \end{enumerate}
\end{enumerate}
\par
On the syntactic side we consider partitions of the set $\Phi$ of formulas into three sets: the set $\Phi_C$ of credible formulas (which contains, at least, all the tautologies), the set $\Phi_A$ of allowable formulas and the set $\Phi_R$ of rejected formulas (which contains, at least, all the contradictions). As in \cite{Bon09}, we use valuations to link syntax and semantics through interpretations and associate, with every interpretation of a GCS, a partial belief revision function. We then define a GCS to be \emph{basic-AGM consistent} if, for every interpretation (or model) of it, the associated partial belief revision function can be extended to a full-domain belief revision function $B_K:\Phi\rightarrow 2^\Phi$ such that, for some basic AGM belief revision function $B^\ast_K:\Phi\rightarrow 2^\Phi$, $\forall\phi\in\Phi$:
\begin{equation*}
B_K(\phi)=
\begin{cases}
K&\text{if } \phi\in\Phi_R\\
B^\ast_K(\phi)&\text{if } \phi\in\Phi_C\\
K\cap B^\ast_K(\phi)&\text{if } \phi\in\Phi_A.
\end{cases}
\end{equation*}
\pagebreak
\par
Thus
\begin{enumerate}
\item if information $\phi$ is rejected then the original beliefs are maintained,
\item if $\phi$ is credible then revision is performed according to the basic AGM postulates, and
\item if $\phi$ is allowable then revision is performed by contracting the original beliefs by the negation of $\phi$ (by the Harper identity the contraction by $\neg\phi$ coincides with taking the intersection of the original beliefs with the revision by $\phi$).
\end{enumerate}
Proposition \ref{Prop2} in Section \ref{Sec:GCS} provides necessary and sufficient conditions for a GCS to be weakly AGM consistent. The proof makes use of the syntactic notion of \emph{filtered belief revision}, which is defined and characterized in Section \ref{Sec:AGM}.
\section{The syntactic approach}
\label{Sec:AGM}
Let $\Phi $ be the set of formulas of a propositional language based on a countable set $A$ of atomic formulas.\footnote{%
Thus $\Phi $ is defined recursively as follows: if $p\in A$ then $p\in \Phi $
and if $\phi ,\psi \in \Phi $ then $\lnot \phi \in \Phi $ and $(\phi \vee
\psi )\in \Phi $.
}
\,Given a subset $K\subseteq \Phi $, its PL-deductive closure $[K]^{PL}$ (where `PL' stands for Propositional Logic) is defined as follows: $\psi \in \lbrack K]^{PL}$ if and only if there exist $\phi _{1},...,\phi _{n}\in K$ \ (with $n\geq 0$) such that $(\phi _{1}\wedge
...\wedge \phi _{n})\rightarrow \psi $ is a tautology (that is, a theorem of Propositional Logic).\footnote{\label{psiinKUphiPL}Note that, if $F$ is a set of formulas, $\psi\in[F\cup\{\phi\}]^{PL}$ if and only if $(\phi\rightarrow\psi)\in[F]^{PL}$.
}\  A set $K\subseteq \Phi $ is \textit{consistent} if $[K]^{PL}\neq \Phi $ (equivalently, if there is no formula $\phi $ such that both $\phi $ and $\lnot \phi $ belong to $[K]^{PL}$).
 A set $K\subseteq \Phi $ is \textit{deductively closed} if $K=\left[ K\right] ^{PL}$. \smallskip

Let $K$ be a consistent and deductively closed set of formulas, representing the agent's initial beliefs, and let $\Psi \subseteq\Phi$  be a set of formulas representing possible items of information.
A \emph{belief revision function based on K and} $\Psi$ is a function $B_{K,\Psi}:\Psi \rightarrow 2^{\Phi }$  that associates with every formula $\phi \in\Psi $ (thought of as new information) a set $B_{K,\Psi}(\phi )\subseteq \Phi $ (thought of as the revised beliefs upon learning that $\phi$). If $\Psi \neq \Phi $ then $B_{K,\Psi}$ is called a \emph{partial} belief revision function, while if $\Psi =\Phi $ then $B_{K,\Psi}$ is called a \emph{full-domain} belief revision function and it is more simply denoted by $B_{K}$. If $B_{K,\Psi}$ is a partial belief revision function and $B'_K$ is a full-domain belief revision function, we say that $B'_K$ is an \emph{extension of} $B_{K,\Psi}$ if, for all $\phi\in\Psi$, $B'_K(\phi)=B_{K,\Psi}(\phi)$.

A \textit{full-domain} belief revision function $B^\ast_K:\Phi\rightarrow 2^\Phi$ is called a \textit{basic AGM function} if it satisfies the first six of the following properties and it is called a \textit{supplemented AGM function} if it satisfies all of them. The following properties are known as the \emph{AGM postulates}:
$\forall \phi ,\psi \in \Phi $,

\begin{center}
\begin{tabular}{ll}
({\small AGM}1) & $B^\ast_K(\phi )=[B^\ast_K(\phi )]^{PL}$\smallskip \\
({\small AGM}2) & $\phi \in B^\ast_K(\phi )$\smallskip \\
({\small AGM}3) & $B^\ast_K(\phi )\subseteq [K\cup\{\phi\} ]^{PL}$\smallskip \\
({\small AGM}4) & if $\lnot \phi \notin K$, then $ [K\cup\{\phi\} ]^{PL}\subseteq B^\ast_{K}(\phi )$\smallskip \\
({\small AGM}5) & $B^\ast_{K}(\phi )=\Phi $ if and only if $\phi $ is a contradiction\smallskip \\
({\small AGM}6) & if $\phi \leftrightarrow \psi $ is a tautology then $B^\ast_{K}(\phi )=B^\ast_{K}(\psi )$ \smallskip\\
({\small AGM}7) & $B^\ast_{K}(\phi \wedge \psi )\subseteq \left[ B^\ast_{K}(\phi )\cup\left\{ \psi \right\} \right] ^{PL}$ \smallskip\\
({\small AGM}8) & if $\lnot \psi \notin B^\ast_{K}(\phi )$, then $\left[B^\ast_{K}(\phi )\cup \left\{ \psi \right\} \right] ^{PL}\subseteq B^\ast_{K}(\phi\wedge \psi ).$\smallskip
\end{tabular}
\smallskip
\end{center}
\par\noindent
{\small AGM}1 requires the revised belief set to be deductively closed.\\[2pt]
{\small AGM}2 postulates that the information be believed. \\[2pt]
{\small AGM}3 says that beliefs should be revised minimally, in the sense that no new formula should be added unless it can be deduced from the information received and the initial beliefs.\footnote{%
Note that (see Footnote \ref{psiinKUphiPL}) $\psi \in \left[ K\cup \{\phi \}\right] ^{PL}$ if
and only if $(\phi \rightarrow \psi )\in K$ (since, by hypothesis, $K=\left[
K\right] ^{PL}$).}\\[2pt]
{\small AGM}4 says that if the information received is compatible with the initial beliefs, then any formula that can be deduced from the information and the initial beliefs should be part of the revised beliefs. \\[2pt]
{\small AGM}5 requires the revised beliefs to be consistent, unless the information $\phi $ is a contradiction (that is, $\lnot \phi $ is a tautology). \\[2pt]
{\small AGM}6 requires that if $\phi $ is propositionally equivalent to $\psi $ then the result of revising by $\phi $ be identical to the result of revising by $\psi $.\smallskip
\par
{\small AGM}1-{\small AGM}6 are called the \emph{basic} AGM postulates, while {\small AGM}7 and {\small AGM}8 are called the \emph{supplementary} AGM postulates.\smallskip
\par\noindent
{\small AGM}7 and {\small AGM}8 are a generalization of {\small AGM}3 and {\small AGM}4 that
\begin{quotation}
\noindent\textquotedblleft applies to \textit{iterated} changes of belief. The idea
is that if $B^\ast_{K}(\phi )$ is a revision of $K$ [prompted by $\phi $] and $%
B^\ast_{K}(\phi )$ is to be changed by adding further sentences, such a change
should be made by using expansions of $B^\ast_{K}(\phi )$ whenever possible.\footnote{%
The expansion of $B^\ast_K(\phi )$ by $\psi $ is $[ B^\ast_{K}(\phi )\cup
\left\{ \psi \right\} ] ^{PL}.$ Note, again, that, for every formula $%
\chi $, $\chi \in [ B^\ast_K(\phi )\cup \left\{ \psi \right\}]
^{PL} $ if and only if $(\psi \rightarrow \chi )\in B^\ast_{K}(\phi )$ (since, by
{\scriptsize AGM1}, $B^\ast_{K}(\phi )=[ B^\ast_{K}(\phi )] ^{PL}$).
}\  More
generally, the minimal change of $K$ to include both $\phi $ and $\psi $
(that is, $B^\ast_{K}(\phi \wedge \psi )$) ought to be the same as the expansion
of $B^\ast_{K}(\phi )$ by $\psi $, so long as $\psi $ does not contradict the
beliefs in $B^\ast_{K}(\phi )$\textquotedblright\ (\cite{Gae88}, p.
55; notation changed to match ours).
\end{quotation}
For an extended discussion of the rationale behind these postulates see \cite{Gae88}.
\par
We now extend the notion of belief revision by allowing the agent to discriminate among different items of information.
\begin{definition}
\label{DEF:PhiPartition}
Let $\Phi$ be the set of formulas of a propositional language. A \emph{credibility partition} is a partition of $\Phi$ into three sets $\Phi_C,\,\Phi_A$ and $\Phi_R$ such that
\begin{enumerate}
\item $\Phi_C$ is the set of \emph{credible} formulas and is such that
\begin{enumerate}
\item it contains all the tautologies,
\item if $\phi\in\Phi_C$ then $\phi$ is consistent,
\item if $\phi\in\Phi_C$ and $\vdash(\phi\leftrightarrow\psi)$ then $\psi\in\Phi_C$, that is, $\Phi_C$ is closed under logical equivalence.
\end{enumerate}
\item $\Phi_A$ is the (possibly empty) set of \emph{allowable} formulas. We assume that if $\phi\in\Phi_A$ then $\phi$ is consistent and that $\Phi_A$ is closed under logical equivalence.
\item $\Phi_R$ is the set of \emph{rejected} formulas, which contains (at least) all the contradictions.
\end{enumerate}
\end{definition}
\begin{definition}
\label{DEF:filtered}
 Let $K$ be a consistent and deductively closed set of formulas (representing the initial beliefs). A (full-domain) belief revision function based on $K$, $B^\circ_{K}:\Phi\rightarrow 2^\Phi$, is called a \emph{filtered belief revision function} if it satisfies the following properties: $\forall \phi,\psi\in\Phi$,
\begin{equation}
\notag
\begin{array}{lll}
 (F1)&\text{if } \phi\in\Phi_{R} &\text{then } B^\circ_{K}(\phi)= K, \smallskip \\
 (F2)&\text{if } \neg\phi\notin K &\text{then } \\
  &&(a)\, \text{ if } \phi\in\Phi_{C} \text{ then }  B^\circ_{K}(\phi)=[ K\cup\{\phi\}]^{PL}\\
  &&(b)\, \text{ if } \phi\in\Phi_{A} \text{ then }  B^\circ_{K}(\phi)= K, \smallskip \\
 (F3)& \text{if  } \neg\phi\in K &\text{then } B^\circ_{K}(\phi)\text{ is consistent and deductively closed and}\\
  &&(a)\, \text{ if } \phi\in\Phi_{C} \text{ then }  \phi\in B^\circ_{K}(\phi)\\
  &&(b)\, \text{ if } \phi\in\Phi_{A} \text{ then }  B^\circ_{K}(\phi)\subseteq (K\setminus \{\neg\phi\})\\
  && \phantom{(b)\, \text{ if } \phi\in\Phi_{A}} \text{ and } [B^\circ_K(\phi)\cup\{\neg\phi\}]^{PL}=K,\smallskip \\
 (F4)&\text{if  } \vdash\phi\leftrightarrow\psi&\text{then } B^\circ_{K}(\phi)=B^\circ_{K}(\psi).
\end{array}
\end{equation}
\end{definition}
\noindent
By ($F1$), if information $\phi$ is rejected, then the original beliefs $K$ are preserved.\\[2pt]
($F2$) says that if, initially, the agent did not believe $\neg\phi$  then  (a) if $\phi$ is credible then the new beliefs are given by the expansion of $K$ by $\phi$, while  (b) if $\phi$ is allowable then the agent does not change her beliefs (since she already considered $\phi$ possible).\\[2pt]
($F3$) says that if, initially, the agent believed $\neg\phi$, then (a) if $\phi$ is credible, then the agent switches from believing $\neg\phi$ to believing $\phi$, while (b) if $\phi$ is allowable, then the agent revises her beliefs by removing $\neg\phi$ from her original beliefs in a minimal way (in the sense that she does not add any new beliefs and if she were to re-introduce $\neg\phi$ into her revised beliefs and close under logical consequence then she would go back to her initial beliefs).\\[2pt]
By ($F4$) belief revision satisfies extensionality: if $\phi$ is logically equivalent to $\psi$ then revision by $\phi$ coincides with revision by $\psi$.
\smallskip\par
The following proposition provides a characterization of filtered belief revision in terms of basic AGM belief revision.\footnote{Note that if $\Phi_A=\varnothing$ then we are in the binary case of ``credibility limited revision'' of \cite{FerHans99,Mak97}.
}
The proof is given in the full paper.

\begin{proposition}
\label{Prop1}
Let $K$ be a consistent and deductively closed set of formulas and $B^\circ_K:\Phi\rightarrow 2^\Phi$ a belief revision function based on $K$. Then the following are equivalent:\\[4pt]
(A) $B^\circ_K$ is a filtered belief revision function,\\[6pt]
(B) there exists a basic AGM belief revision function $B^\ast_K:\Phi\rightarrow 2^\Phi$ such that,  $\forall\phi\in\Phi$,
\begin{equation}
\label{filteredAGM}
B^\circ_K(\phi)=
\begin{cases}
K &\quad \text{if } \phi\in\Phi_R\\
B^\ast_K(\phi) &\quad\text{if } \phi\in\Phi_C\\
K\cap B^\ast_K(\phi)  &\quad\text{if } \phi\in\Phi_A.
\end{cases}
\end{equation}
\end{proposition}
\noindent
(\ref{filteredAGM}) says the following:
\begin{enumerate}
\item if information $\phi$ is rejected then the original beliefs are maintained,
\item if $\phi$ is credible then revision is performed according to the AGM postulates, and
\item if $\phi$ is allowable then revision is performed by contracting the original beliefs by the negation of $\phi$ (by the Harper identity the contraction by $\neg\phi$ coincides with taking the intersection of the original beliefs with the revision by $\phi$).
\end{enumerate}
Note that if $\neg\phi\notin K$ then $B^\ast_K(\phi)=\left[K\cup\{\phi\}\right]^{PL}\supseteq K$  and thus $K\cap B^\ast_K(\phi)=K$ so that information $\phi\in\Phi_A$ has no effect on the initial beliefs. Thus,  if $\phi\in\Phi_A$, belief change occurs only when $\neg\phi\in K$, that is, when -- initially -- the agent believes $\neg \phi$; in this case, since $\neg\phi\in K$ (implying, by consistency of $K$, that $\phi\notin K$) and $\phi\in B^\ast_K(\phi)$ (implying, by consistency of $B^\ast_K(\phi)$, that $\neg\phi\notin B^\ast_K(\phi)$), it follows that $\phi\notin B^\circ_K(\phi)$ and $\neg\phi\notin B^\circ_K(\phi)$, so that the agent's reaction to being informed that $\phi$ (with $\phi\in\Phi_A$)  is to suspend judgment concerning $\phi$, in other words, to consider both $\phi$ and $\neg\phi$ as possibilities.\medskip\par

\section{Generalized choice structures}
\label{Sec:GCS}
\begin{definition}
\label{Def:GCS}
A \emph{generalized choice structure} (GCS) is a tuple $\left\langle \Omega ,\{\mathcal{E}_C, \mathcal E_A, \mathcal E_R\}, f\right\rangle$ such that:
\begin{enumerate}
  \item $\Omega\neq\varnothing$,
  \item $\mathcal{E}_C, \mathcal E_A, \mathcal E_R$ are mutually disjoint subsets of $2^\Omega$ with $\Omega\in\mathcal{E}_C$ and $\varnothing\in\mathcal{E}_R$,
  \item $f:\mathcal E\rightarrow 2^\Omega$ (where $\mathcal E=\mathcal{E}_C\cup \mathcal E_A\cup \mathcal E_R$) is such that
  \begin{enumerate}
    \item $f(\Omega)\neq\varnothing$,
    \item if $E\in\mathcal E_R$ then $f(E)=f(\Omega)$,
    \item if $E\in\mathcal E_C$ then $\varnothing\neq f(E)\subseteq E$,
    \item if $E\in\mathcal E_A$ then $f(E)\cap E\neq\varnothing$.
  \end{enumerate}
\end{enumerate}
\end{definition}
Next we introduce the notion of a model, or interpretation, of a GCS.
\par
Fix a propositional language based on a countable set $A$ of atomic formulas and let $\Phi$ be set of formulas.
A \textit{valuation} is a function $V:A\rightarrow 2^{\Omega }$ that associates with every atomic
formula $p\in A$ the set of states at which $p$ is true. Truth of an arbitrary formula at a state is defined
recursively as follows ($\omega \models\phi $ means that
formula $\phi $ is true at state $\omega $): \\
(1) for $p\in A$, $\omega \models p$ if and only if $\omega
\in V(p)$, \\
(2) $\omega \models\lnot \phi $ if and only if $\omega \not\models\phi $ and\\
(3) $\omega \models (\phi \vee \psi )$ if and only if either $\omega \models \phi$ or $\omega \models \psi$ (or both). \\
The truth set of formula $\phi$ is denoted by $||\phi
||$. Thus $|| \phi ||=\{\omega \in \Omega :\omega \models \phi \}$.
\par
Given a valuation $V$, define:\footnote{
All these objects, including the truth sets of formulas, are dependent on the valuation $V$ and thus ought to be indexed by it; however, in order to keep the notation simple, we will not add the subscript $`V$' to each of these objects.
}
\begin{gather}
\label{KV}
K=\left\{\phi\in\Phi:f(\Omega)\subseteq ||\phi||\right\},\\
\label{PsiV}
\Psi=\left\{\phi\in\Phi:||\phi||\in\mathcal E\right\}\text{ and}\\
\label{BV}
B_{K,\Psi}:\Psi\to 2^\Phi\text{ given by: } B_{K,\Psi}(\phi)=\left\{\psi\in\Phi:f\left(||\phi||\right)\subseteq ||\psi||\right\}.
\end{gather}
Since $f(\Omega)$ is interpreted as the set of states that the individual initially considers possible, (\ref{KV}) is the initial belief set. It is straightforward to show that $K$ is consistent (since $f(\Omega)\neq\varnothing$) and deductively closed.\\[2pt]
(\ref{PsiV}) is the set of formulas that are potential items of information.\\[2pt]
(\ref{BV}) is the partial belief revision function encoding the agent's disposition to revise her beliefs in response the items of information in $\Psi$ (for $E\in\mathcal E$, $f(E)$ is interpreted as the set of states that the individual considers possible after receiving information $E$).

\begin{definition}
\label{Def:Model}
Given a GCS $\left\langle \Omega ,\{\mathcal{E}_C, \mathcal E_A, \mathcal E_R\}, f\right\rangle$, a \emph{model} or \emph{interpretation} of it is obtained by adding to it a pair $\left(\left\{\Phi_C,\Phi_A,\Phi_R\right\},V\right)$, where
$\left\{\Phi_C,\Phi_A,\Phi_R\right\}$ is a credibility partition of $\Phi$ (Definition \ref{DEF:PhiPartition}) and $V$ is a valuation such that, $\forall\phi\in\Phi$,
\begin{enumerate}
\item if $||\phi||\in\mathcal E_C$ then $\phi\in\Phi_C$,
\item if $||\phi||\in\mathcal E_A$ then $\phi\in\Phi_A$,
\item if $||\phi||\in\mathcal E_R$ then $\phi\in\Phi_R$.
\end{enumerate}
\end{definition}
\begin{definition}
\label{Def:basicAGMconsistent}
A generalized choice structure $\mathcal C=\left\langle \Omega ,\{\mathcal{E}_C, \mathcal E_A, \mathcal E_R\}, f\right\rangle$ is \emph{basic-AGM consistent} if, for every model $\left\langle\mathcal C,\left\{\Phi_C,\Phi_A,\Phi_R\right\},V\right\rangle$ of it, letting $B_{K,\Psi}$ be the corresponding partial belief revision function (defined in (\ref{BV})), there exist
\begin{enumerate}
\item a full-domain belief revision function $B^\circ_K:\Phi\to 2^\Phi$ that extends $B_{K,\Psi}$ (that is, for every $\phi\in\Psi$, $B^\circ_K(\phi)=B_{K,\Psi}(\phi)$) and
\item a basic AGM belief revision function $B^\ast_K:\Phi\to 2^\Phi$
\end{enumerate}
such that, for every $\phi\in\Phi$
\begin{equation}
\label{filtered AGM}
B^\circ_K(\phi)=
\begin{cases}
K &\quad \text{if } \phi\in\Phi_R\\
B^\ast_K(\phi) &\quad\text{if } \phi\in\Phi_C\\
K\cap B^\ast_K(\phi)  &\quad\text{if } \phi\in\Phi_A.
\end{cases}
\end{equation}
\end{definition}
The following proposition gives necessary and sufficient conditions for a GCS to be basic-AGM consistent. The proof is given in the full paper.
\begin{proposition}
\label{Prop2}
Let $\mathcal C=\left\langle \Omega ,\{\mathcal{E}_C, \mathcal E_A, \mathcal E_R\}, f\right\rangle$ be a generalized choice structure. Then the following are equivalent:\\[6pt]
(A) $\mathcal C$ is basic-AGM consistent\\[8pt]
(B) $\mathcal C$ satisfies the following properties: for every $E\in\mathcal E$,
\begin{enumerate}
\item if $E\cap f(\Omega)\neq\varnothing$ then
\begin{enumerate}
\item if $E\in\mathcal E_C$ then $f(E)=E\cap f(\Omega)$,
\item $E\in\mathcal E_A$ then $f(E)=f(\Omega)$,
\end{enumerate}
\item if $E\cap f(\Omega)=\varnothing$ and $E\in\mathcal E_A$ then $f(E)=f(\Omega)\cup E'$ for some $\varnothing\neq E'\subseteq E$.
\end{enumerate}
\end{proposition}
\vspace{10pt}
\par
In the full paper we also define the notion of a rationalizable GCS and establish a correspondence between the notion of generalizable GCS and the notion of strong AGM consistency.
\newpage

\nocite{*}
\bibliographystyle{eptcs}
\bibliography{TARK}
\end{document}